\documentclass[aps, prb, twocolumn, amssymb, amsmath, float]{revtex4-1}

\usepackage{graphicx,dcolumn,bm,hyperref,subfigure,multirow, color}
\begin{document}

\title{Pressure-stabilized binary compounds of magnesium and silicon}

\author{Tran Doan Huan}%
\affiliation{Department of Materials Science and Engineering and Institute of Materials Science, University of Connecticut, Storrs, CT 06269, USA}
\email[Email:]{huan.tran@uconn.edu}
\date{\today}

\begin{abstract}
The family of binary compounds composed of magnesium and silicon is rather rich. In addition to the well-known magnesium silicide Mg$_2$Si, other compounds, including MgSi$_2$, Mg$_4$Si$_7$, Mg$_5$Si$_6$, MgSi, and Mg$_9$Si$_5$, have also been identified and/or proposed in precipitated Al-Mg-Si solid solutions. Nevertheless, computational studies show that only Mg$_2$Si is thermodynamically stable at ambient conditions while certain non-zero hydrostatic pressure can stabilize Mg$_9$Si$_5$ so that it can co-exist with Mg$_2$Si. We conduct a comprehensive search for viable binary compounds of Mg$_x$Si$_{1-x}$ ($1/3\leq x \leq 2/3$), discovering numerous new structures for all the compounds. On the one hand, we find that MgSi$_2$, MgSi, and Mg$_9$Si$_5$ are likely pressure-stabilized materials, while, on the other hand, supporting previous studies, raising doubt on the existence of Mg$_5$Si$_6$, and claiming that the existence of Mg$_4$Si$_7$ remains an open question. Therefore, we recommend that (hydrostatic and/or non-hydrostatic) pressure should be explicitly considered when discussing the stability of these solids (and maybe other solids as well) by computations. We also find that MgSi$_2$ can potentially exhibit superconducting behaviors within a wide range of pressure with the critical temperature of up to $7$ K.
\end{abstract}

\maketitle

\section{Introduction}
The best-known binary compound of Mg and Si, i.e., magnesium silicide Mg$_2$Si, has been studied extensively due to its potential applications, e.g., infrared photonic and thermoelectric energy conversion devices.\cite{Udono2015, Udono2013, Zaitsev2006, Liu2012, Morozova2014} Five other members of the Mg-Si binary family, including MgSi$_2$, Mg$_4$Si$_7$, Mg$_5$Si$_6$, MgSi, and Mg$_9$Si$_5$, have subsequently been identified/suggested experimentally (and occasionally studied computationally) when precipitation-hardened alloys of Al with Mg and Si were explored.\cite{vanHuis20062945, VANHUIS20072183, Zhang2015, Vissers20073815, Ji:Mg2Si, ZandbergenScience, Andersen19983283, Matsuda1998, Matsuda2000, ANDERSEN2005127} Of them, the $P4/mmm$ phase of MgSi and the $P6_3/m$ phase of Mg$_9$Si$_5$ have been confirmed experimentally\cite{Matsuda1998, Jacobs:Mg9Si5} and studied computationally.\cite{Ravi20044213, Vissers20073815} The others, e.g., MgSi$_2$, Mg$_4$Si$_7$, and Mg$_5$Si$_6$ were proposed computationally as candidates for many (still) unknown phases of the Al-Mg-Si alloys. Specifically, Mg$_5$Si$_6$, assumed to be in a $C2/m$ monoclinic phase, was initially proposed \cite{ZandbergenScience, 0953-8984-14-15-315} as the $\beta''$ precipitate of the Al-Mg-Si alloys but this proposal has then caught considerable doubt.\cite{HastingJAP106, EHLERS2014617, qiu_kong_xiao_du_2016, Ninive2014, FLAMENT201764} Nevertheless, the observation of these Mg-Si binary compounds seems to contradict some first-principles calculations performed at zero pressure ($P=0$ GPa),\cite{Ravi20044213, vanHuis20062945, OpahleMgSi} revealing that only Mg$_2$Si is thermodynamically stable, while other Mg-Si binary compounds are unstable.

Among these binary compounds, Mg$_5$Si$_6$ and Mg$_9$Si$_5$ were recently predicted\cite{Huan:Mg2Si} to become stable under certain ranges of compressive hydrostatic pressure. This computational study suggests the possible role of pressure, the thermodynamic variable that may be realized in certain experimental conditions, in the observations of Mg$_5$Si$_6$, Mg$_9$Si$_5$, and possibly other Mg-Si binary compounds as well.\cite{Huan:Mg2Si} In fact, pressure has already been known as a key factor, stabilizing numerous new solid materials with exotic functionalities, e.g., high energy density and high-temperature superconductivity,\cite{Zhang1502, Drozdov15, Alexey:CaB, ANGE:ANGE201701660, C6CP08708F, Yu:SnSe} and inspiring a great deal of interest from the community.\cite{Tuan:PSSB, C4CP03207A, Ma_perspect, Zhang:NRM17, Eva:highpressure}

Compared to Mg$_2$Si, not much was known \cite{Ravi20044213, Huan:Mg2Si, Huan:Mg5Si6} about MgSi, MgSi$_2$, Mg$_4$Si$_7$, Mg$_5$Si$_6$, Mg$_9$Si$_5$, and other possible binary compounds of this family. Recently, the thermodynamic stability and the electronic structure of Mg$_5$Si$_6$ and Mg$_9$Si$_5$ (whose the hexagonal $P6_3/m$ structure was proposed\cite{Jacobs:Mg9Si5,Vissers20073815,Ravi20044213} as the $\beta'$ precipitate of the Al-Mg-Si alloy systems) have been exploited by first-principles computations.\cite{Ravi20044213, OpahleMgSi, Huan:Mg2Si, Huan:Mg5Si6} Under compressive pressure ($P \gtrsim 10$ GPa and above), they were predicted\cite{Huan:Mg5Si6} to be dynamically stable and share the metallic/semimetallic characteristics with Mg$_2$Si. To our best knowledge, an in-depth understanding of the other binary compounds, if realized, remains unavailable.

This contribution addresses these two points. By searching for low-energy structures of 13 Mg-Si binary compounds (with varying Mg content) at the level of density functional theory (DFT), \cite{DFT1, DFT2} we identified numerous structures that are significantly lower in energy than those currently recognized. Because the number of Mg-Si binary compounds considered in this comprehensive work is sufficiently large, the stability of these compounds can be better accessed. From this analysis, we suggest that MgSi$_2$, MgSi, and Mg$_9$Si$_5$ are likely pressure-stabilized materials while confirming (and suggesting) the thermodynamical instability of Mg$_5$Si$_6$\cite{HastingJAP106, EHLERS2014617, qiu_kong_xiao_du_2016, Ninive2014, FLAMENT201764} (and Mg$_4$Si$_7$). We recommend to explicitly consider external pressure when discussing the thermodynamical stability of Mg-Si based solids, and presumably other solids as well. For those identified to be thermodynamically stable, their dynamical stability, electronic structure, and possible superconductivity were studied using DFT computations.

In addition to the aforementioned results, the dataset of 358 low-energy structures identified herein is also useful for the community in the context of the emerging age of materials informatics.\cite{Rajan2005, Huan:design, Arun:design} Because this dataset was prepared by exhaust low-energy structure searches, it provides a large number of new stable materials structures. Generally, datasets prepared in this way\cite{Huan:Data, Chiho:hybrid} are good complement to the established materials databases such as Materials Project,\cite{MatProj} Open Quantum Materials Database,\cite{OQMD} AFLOW,\cite{AFLOWLIB} and Polymer Genome.\cite{MannodiKanakkithodi2017}

\section{Computational details}
Our first-principles calculations were performed at the level of density functional theory, using specifically the version implemented in {\it Vienna Ab initio Simulation Package}. \cite{vasp1,vasp2,vasp3,vasp4} We used the kinetic energy cutoff of 500 eV for the plane-wave basis set, and the generalized gradient approximation Perdew-Burke-Ernzerhof functional\cite{PBE} for the exchange-correlation (XC) energies. The Brillouin zone of the examined structures was sampled by a Monkhorst-Pack $k$-point mesh\cite{monkhorst} with a spacing of 0.1 \AA$^{-1}$ in the reciprocal space.

\begin{table}[t]
\caption{Summary of the Mg-Si binary compounds on which the searches were performed in this work. For each binary, the Mg concentration $x$ and $N_{\rm max}$, the maximum number of atoms of the cells used for the search, are given.}\label{table:sum}
\begin{center}
\begin{tabular}{c c c c c}
\hline
\hline
Materials  &  $x$ & $N_{\rm max}$ & No. f.u. & No. structs. \\
\hline
MgSi$_2$     & $0.333$            & 24& 8  & 65 \\
Mg$_5$Si$_9$ & $0.357$            & 28& 2  & 14 \\
Mg$_4$Si$_7$ & $0.364$            & 22& 2  & 26 \\
Mg$_2$Si$_3$ & $0.400$            & 20& 4  & 29 \\
Mg$_3$Si$_4$ & $0.429$            & 28& 4  & 43 \\
Mg$_5$Si$_6$ & $0.455$            & 22& 2  & 32 \\
MgSi         & $0.500$            & 24& 12 & 34 \\
Mg$_6$Si$_5$ & $0.545$            & 22& 2  & 15 \\
Mg$_4$Si$_3$ & $0.571$            & 28& 4  & 41 \\
Mg$_3$Si$_2$ & $0.600$            & 20& 4  & 9 \\
Mg$_7$Si$_4$ & $0.636$            & 22& 2  & 15\\
Mg$_9$Si$_5$ & $0.643$            & 28& 2  & 14\\
Mg$_2$Si     & $0.667$            & 24& 8  & 21 \\
\hline
\end{tabular}
\end{center}
\end{table}

Low-energy structures of the possible Mg-Si binary compounds was searched using the minima-hopping method,\cite{Goedecker:MHM, Amsler:MHM, MHM:OrganovBookChapter} which has been successfully used for different material classes \cite{Huan:Zn, Huan:Mixed, Huan:NaSc, Huan:Mg2Si, Huan:perovskites} This method relies on exploring the DFT energy landscape by alternating molecular-dynamics runs for escaping the current local minimum and geometry optimization runs for identifying the next local minimum. As the energy landscapes of the examined materials are constructed at the DFT level, the searches are reliable but generally expensive. In principles, searches can be performed at any pressure, as performed in Refs. \onlinecite{Huan:Mg2Si}; however, given the number of binary compounds considered is large (13) and that the pressure window of interest is unknown, the searches for each Mg-Si binary were conducted only at zero pressure for some certain numbers of formula units (and equivalently, number of atoms --- see Table \ref{table:sum} for details). The structures within a windows of $200$ meV/atom from the lowest-energy structure were selected and studied at varying elevated pressures. This procedure, as will subsequently be shown in this work, captures the stable stoichiometries of the Mg$_x$Si$_{1-x}$ identified within $0-30$ GPa, the pressure range of our current interest. Their symmetry was analyzed using {\sc findsym}\cite{findsym} while {\sc vesta}\cite{vesta} was used for visualization. The (crystallographic information format) structure files were prepared with {\sc pymatgen}.\cite{pymatgen}

\begin{table}[t]
\caption{Summary of the new structures predicted for Mg-Si binary compounds. For those predicted in this work, $\Delta H_{\rm DFT}$ is the enthalpy of formation, given in meV/atom.}\label{table:new}
\begin{center}
\begin{tabular}{c c c c c c c }
\hline
\hline
\multirow{2}{*}{Mater.}   & &  \multicolumn{2}{c}{Literature} & & \multicolumn{2}{c}{This work} \\
\cline{3-4} \cline{6-7}
& & Symmetry & Refs. && Symmetry & $\Delta H_{\rm DFT}$ \\
\hline
\multicolumn{7}{c}{$P=0$ GPa}\\
	MgSi$_2$     && $Imma$           & \onlinecite{vanHuis20062945} & & $R\overline{3}m$& $-90.2$  \\
	Mg$_4$Si$_7$ && $C2/m$           & \onlinecite{vanHuis20062945, VANHUIS20072183}&& $P1$  & $-6.1$ \\
	Mg$_5$Si$_6$ && $C2/m$           & \onlinecite{ZandbergenScience,Andersen19983283}&& $C2/m$ & $-5.3$ \\
	MgSi         && $P4/mmm$         & \onlinecite{Matsuda1998,Ravi20044213}&& $P2_1/m$ & $-140$ \\
	Mg$_9$Si$_5$ && $P6_3/m$         & \onlinecite{Jacobs:Mg9Si5,Vissers20073815}&& $R\overline{3}c$  & $-3.0$\\
\hline
\multicolumn{7}{c}{$P=10$ GPa}\\
	MgSi$_2$     && $Imma$            & \onlinecite{vanHuis20062945} & & $Imma$ & $-1.1$  \\
	Mg$_4$Si$_7$ && $C2/m$            & \onlinecite{vanHuis20062945, VANHUIS20072183}&& $P1$ & $-6.2$ \\
	Mg$_5$Si$_6$ && $C2/m$            & \onlinecite{ZandbergenScience,Andersen19983283}&& $C2/m$  & $-0.8$ \\
	MgSi         && $P4/mmm$          & \onlinecite{Matsuda1998,Ravi20044213}&& $C2/m$ & $-38.5$ \\
	Mg$_9$Si$_5$ && $P6_3/m$          & \onlinecite{Jacobs:Mg9Si5,Vissers20073815}&& $R\overline{3}c$  & $-0.8$ \\
\hline
\multicolumn{7}{c}{$P=20$ GPa}\\
	MgSi$_2$     && $Imma$            & \onlinecite{vanHuis20062945}      & & $P6/mmm$ & $ -3.2$  \\
	Mg$_4$Si$_7$ && $C2/m$            & \onlinecite{vanHuis20062945, VANHUIS20072183}      & & $Pm$   & $-28.1$ \\
	Mg$_5$Si$_6$ && $C2/m$            & \onlinecite{ZandbergenScience,Andersen19983283}&& $Cm$ & $-5.2$ \\
	MgSi         && $P4/mmm$          & \onlinecite{Matsuda1998,Ravi20044213}&& $P4/mmm$  & $0$\\
	Mg$_9$Si$_5$ && $P6_3/m$          & \onlinecite{Jacobs:Mg9Si5,Vissers20073815}&& $R\overline{3}c$ & $-1.0$ \\
\hline
\multicolumn{7}{c}{$P=30$ GPa}\\
	MgSi$_2$     && $Imma$            & \onlinecite{vanHuis20062945} & & $P6/mmm$ & $-6.7$  \\
	Mg$_4$Si$_7$ && $C2/m$            & \onlinecite{vanHuis20062945, VANHUIS20072183}&& $Cm$  & $-107$ \\
	Mg$_5$Si$_6$ && $C2/m$            & \onlinecite{ZandbergenScience,Andersen19983283}&& $Cm$  & $-46.7$ \\
	MgSi         && $P4/mmm$          & \onlinecite{Matsuda1998,Ravi20044213}&& $P4/mmm$ & $0$\\
	Mg$_9$Si$_5$ && $P6_3/m$          & \onlinecite{Jacobs:Mg9Si5,Vissers20073815}&& $R\overline{3}c$ & $-4.3$ \\
\hline
\end{tabular}
\end{center}
\end{table}

\begin{figure*}[t]
  \begin{center}
    \includegraphics[width= 17.5 cm]{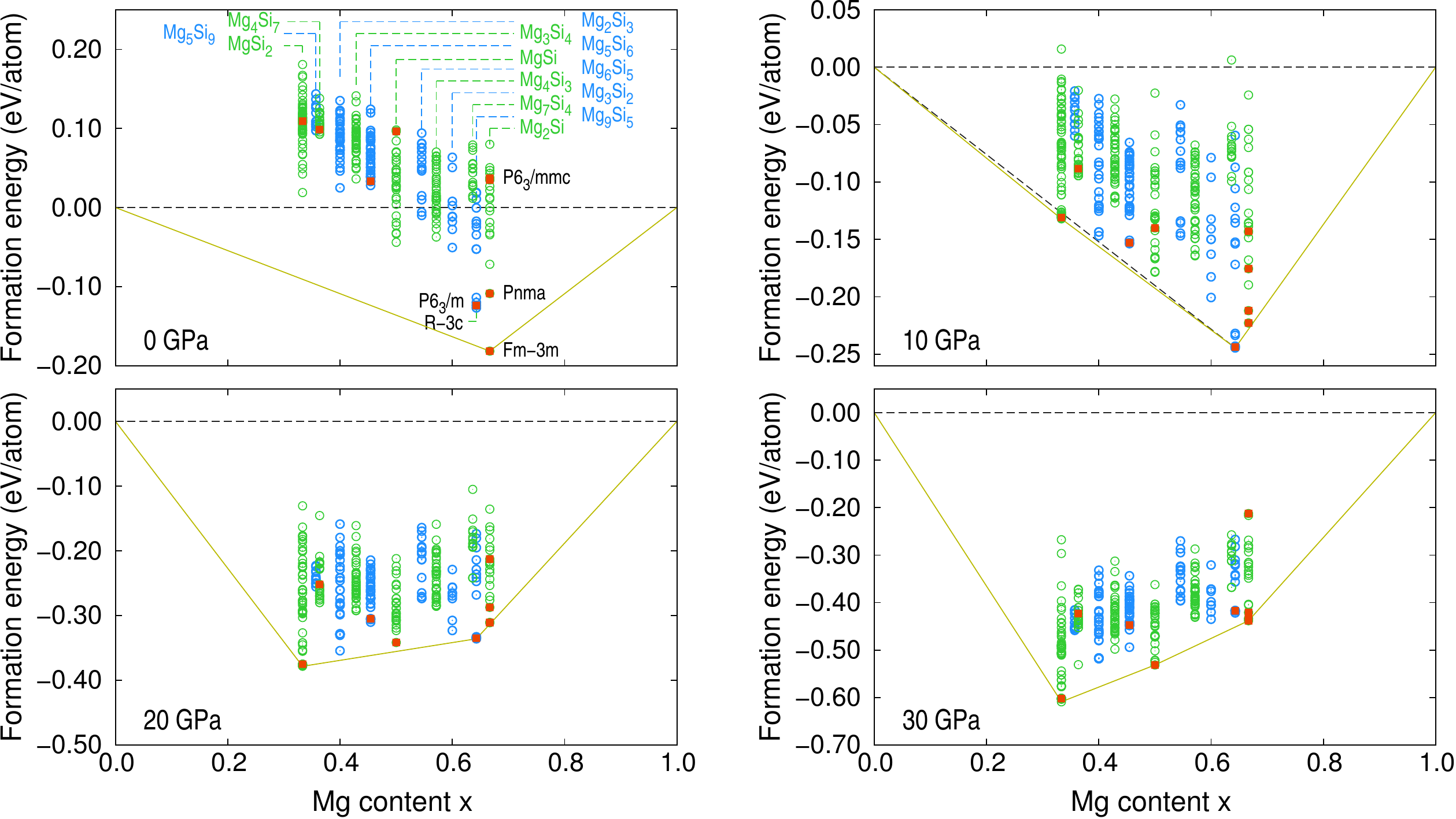}
  \caption{(Color online) Convex hull constructed from the DFT enthalpy of formation $\Delta H_{\rm DFT}$  defined in Eq. \ref{eq:fme} for Mg$_x$Si$_{1-x}$ binary compounds. Alternative colors are used to label the binary compounds while red squares are used for the structures obtained either experimentally or computationally in the literature.} \label{fig:hull}
  \end{center}
\end{figure*}

Phonon-related calculations reported herein, including those related to the phonon-mediated superconductivity of the predicted structures as discussed subsequently, were performed using the linear response approach\cite{Abinit_phonon_2,Abinit_phonon_3} as implemented in {\sc abinit}, \cite{Gonze_Abinit_3} employing the Hartwigsen-Goedecker-Hutter norm-conserving pseudopotentials,\cite{HGH_Pseudo} a plane-wave cutoff energy of 40 Hatree ($\simeq1,100$eV) and the PBE XC functional. In fact, because of some intractable uncertainty, material structures determined computationally or experimentally may actually be dynamically unstable and in this case, proper phonon calculations may be used to refine them.\cite{Huan:NaSc,C6TA02761J} The phonon band structure reported in this work are used to track the dynamical stability of the structure predicted. Correction to the energy from the lattice vibrations, which can be computed from the phonon density of states (as performed in Refs. \onlinecite{Huan:NaSc,Huan:Mg2Si,Huan:perovskites}), was not considered here because of two reasons. First, the computational resource required for more than three hundred structures at different values of pressure is prohibitively enormous, and second, zero-point energy corrections tend to cancel out, leaving only a part of such small amount contributing to the relative energy difference between structures.

\section{Results and discussions}
\subsection{Thermodynamic stability}
A summary of the low-energy structures identified in this work is given in Table \ref{table:new} while their detailed cystallographic information is provided in the Supplemental Materials.\cite{supplement} Except Mg$_2$Si, new ``ground state'' structures were identified for the others at $P=0$ GPa. For MgSi$_2$ and MgSi, the new lowest-energy structures ($R\overline{3}m$ and $P2_1/m$) are significantly lower than the previously reported counterparts in $E_{\rm DFT}$ by $\simeq 90$ meV/atom and $\simeq 140$ meV/atom, respectively. For Mg$_4$Si$_7$, Mg$_5$Si$_6$, and Mg$_9$Si$_5$, the advance in $E_{\rm DFT}$ of the new structures is smaller but remains noticeable.

The thermodynamic stability of the identified structures are examined by four convex hulls shown in Fig. \ref{fig:hull}. They were constructed at $P=0$, $10$, $20$, and $30$ GPa from the formation DFT enthalpy $\Delta H_{\rm DFT}$, defined as
\begin{eqnarray} \label{eq:fme}
\Delta H_{\rm DFT} = H_{\rm DFT}({\rm Mg}_x{\rm Si}_{1-x}) - \left[xH_{\rm DFT}({\rm Mg})\right.\nonumber \\
\left. + (1-x)H_{\rm DFT}({\rm Si})\right].
\end{eqnarray}
Here, $H_{\rm DFT}({\rm Mg}_x{\rm Si}_{1-x})$, $H_{\rm DFT}({\rm Mg})$, and $H_{\rm DFT}({\rm Si})$ are the DFT enthalpies computed for ${\rm Mg}_x{\rm Si}_{1-x}$, the ground state hexagonal $P6_3/mmc$ structure of Mg, and the ground state cubic $Fd\overline{3}m$ structure of Si. In the definition $H_{\rm DFT}\equiv E_{\rm DFT} + PV$ of the DFT enthalpy, the DFT energy $E_{\rm DFT}$ and the volume of the simulation box $V$ were computed at the hydrostatic pressure $P$.

\begin{figure*}[t]
\begin{center}
\includegraphics[width= 15.0 cm]{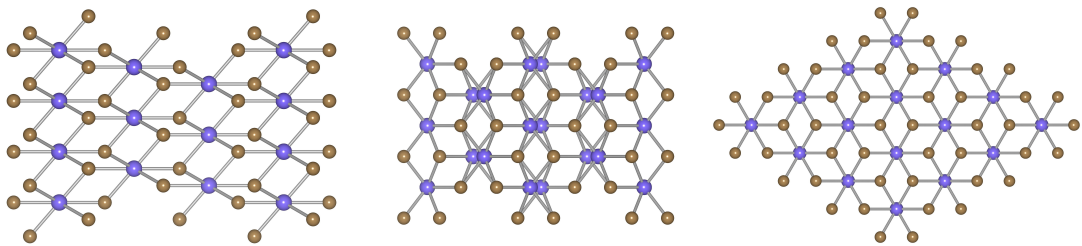}
\caption{(Color online) Structures of MgSi$_2$ at 0 GPa ($R\overline{3}m$), 10 GPa ($Imma$), and 20 GPa ($P6/mmm$). The structure at 30 GPa is similar to that at 20 GPa. Magnesium and silicon atoms are shown in olive and medium slate blue colors, respectively.} \label{fig:strMgSi2}
\end{center}
\end{figure*}

In consistence with previous reports,\cite{Ravi20044213, vanHuis20062945, OpahleMgSi, Huan:Mg2Si} only Mg$_2$Si is thermodynamically stable at $P=0$ GPa. Starting from $P\simeq 10$ GPa, MgSi$_2$ becomes stable in different phases, i.e., $Imma$ and $P6/mmm$ at $P=10$ and $\geq 20$ GPa, respectively. The predicted $Imma$ structure (with $a=4.12$ \AA, $b=5.64$ \AA, $c=7.64$ \AA) is about 1 meV/atom lower than the previously proposed $Imma$ structure \cite{vanHuis20062945} (with $a=4.00$ \AA, $b=5.88$ \AA, $c=7.60$ \AA) but they appear to be just slightly different when the motifs are visualized. At 20 GPa and 30 GPa, the predicted $P6/mmm$ structure of MgSi$_2$ is new, as shown in Fig. \ref{fig:strMgSi2}. Similarly, the predicted $R\overline{3}c$ structure of Mg$_9$Si$_5$ (summarized in Table \ref{table:new}) is lower than the $P6_3/m$ structure previously proposed\cite{Jacobs:Mg9Si5, Vissers20073815} (and studied computationally\cite{Huan:Mg5Si6}) by $\simeq 1-4$ meV/atom but a closer investigation indicates that the difference between them is also mall (see Supplemental Materials\cite{supplement} for a visualization). Considering the small energy difference, the recent conclusion that between $6$ and $24$ GPa, Mg$_2$Si can decompose into Mg$_9$Si$_5$ and Mg without energy cost\cite{Huan:Mg2Si} remains valid, as shown in Fig. \ref{fig:hull}. For MgSi, the previously proposed $P4/mmm$ structure \cite{Matsuda1998,Ravi20044213} is higher than the predicted $P2_1/m$ structure by $\simeq 140$ meV/atom at $P=0$ GPa. However, this shortcoming is rapidly diminished as $P$ increases, and starting from $P \simeq 20$ GPa, the $P4/mmm$ structure of MgSi becomes lowest in $H_{\rm DFT}$.

The above observation strongly hints that MgSi$_2$\cite{vanHuis20062945} and specifically MgSi\cite{Matsuda1998, Ravi20044213} and Mg$_9$Si$_5$,\cite{Jacobs:Mg9Si5, Vissers20073815} whose experimentally observed structures become lowest in enthalpy within some ranges of pressure, are indeed pressure-stabilized materials. From the computational point of view, (hydrostatic and/or non-hydrostatics) pressure should be explicitly considered when discussing the thermodynamic stability of the Mg-Si binary compounds. This conclusion, which places MgSi$_2$, MgSi, and Mg$_9$Si$_5$ into a class of (pressure-stabilized) metastable materials like LiK(BH$_4$)$_2$,\cite{Tuan:PSSB, C4CP03207A} aligns well with the recent rising role of pressure that has been extensively discussed in the literature of materials discovery.\cite{Ma_perspect, Zhang:NRM17, Eva:highpressure}

On the other hand, Table \ref{table:new} and Fig. \ref{fig:hull} show that Mg$_4$Si$_7$ and Mg$_5$Si$_6$, which have also been reported previously,\cite{vanHuis20062945, VANHUIS20072183, ZandbergenScience, Andersen19983283} are thermodynamically unstable at pressure $P$ up to 30 GPa. For both of them, their lowest-enthalpy structure are always about $100-200$ meV/atom above the convex hulls. This observation is consistent with more recent findings\cite{HastingJAP106, EHLERS2014617, qiu_kong_xiao_du_2016, Ninive2014,FLAMENT201764} that the $\beta''$ phase of Al-Mg-Si is not Mg$_5$Si$_6$ as previously conjectured. In case of Mg$_4$Si$_7$, the theoretically proposed structures\cite{vanHuis20062945,Zhang2015} were shown\cite{Zhang2015} to have positive formation energy at 0 GPa. These results, together with what revealed by Fig. \ref{fig:hull} that $\Delta H_{\rm DFT}$ of Mg$_4$Si$_7$ is significantly higher than the convex hulls at any $P$, suggest that the existence of Mg$_4$Si$_7$ is an open question.

Among the other binary compounds examined, Mg$_2$Si$_3$ and Mg$_3$Si$_2$ are ``nearly'' stable within $10-20$ GPa, where their lowest-enthalpy structures are just about $1-5$ meV/atom about the convex hulls. Although no direct report for these compounds are currently available, there are however a fair number of phases of the Al-Mg-Si alloys that have yet been resolved.\cite{Ravi20044213} As summarized by Table \ref{table:new} and Fig. \ref{fig:hull}, pressure strongly alters the energetic ordering of the low-lying structures, being an important factor leading to the significant complexity of the systems.

\subsection{MgSi$_2$}\label{sec:MgSi2}
\begin{figure}[t]
\begin{center}
\includegraphics[width= 8.25 cm]{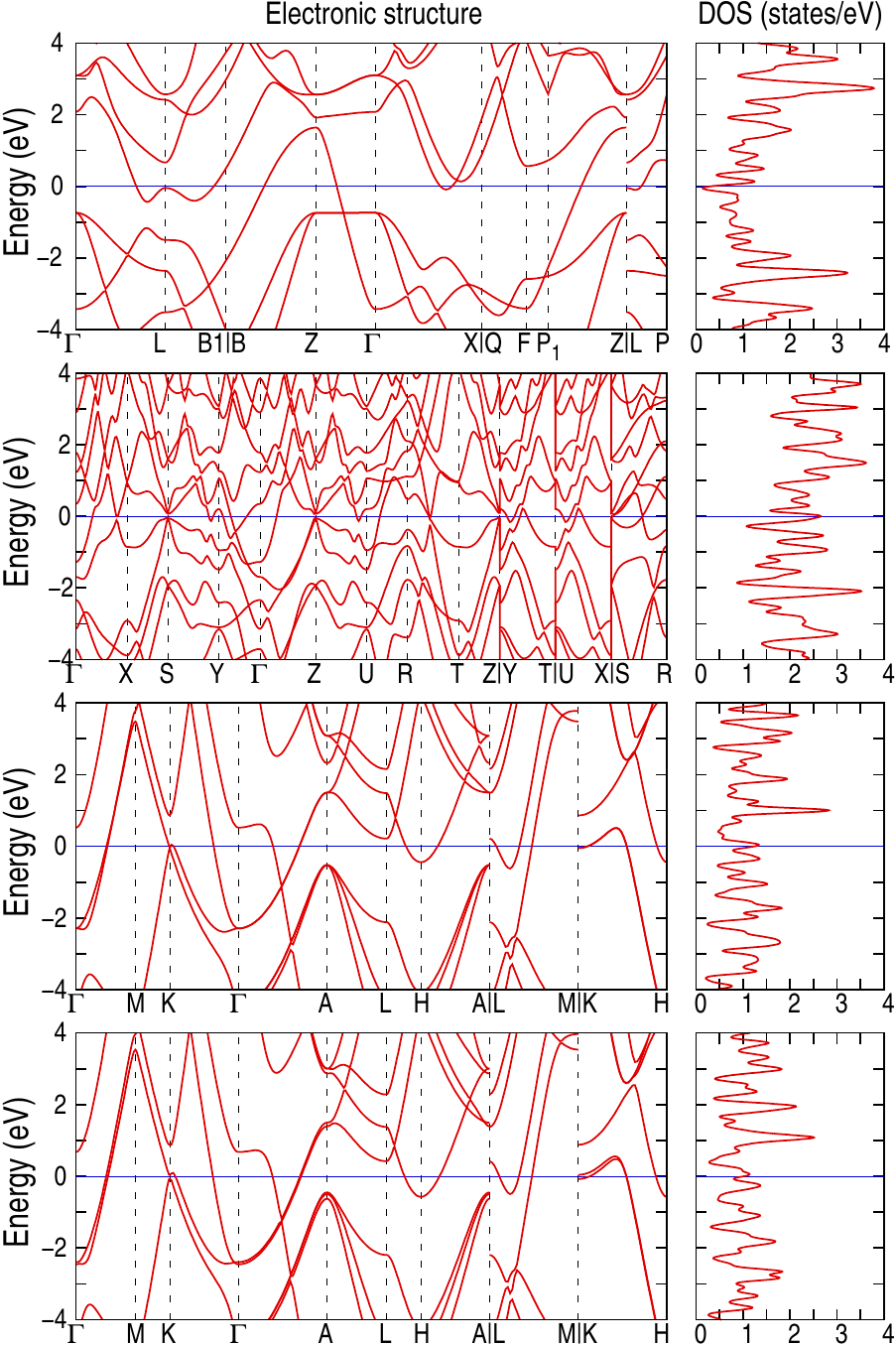}
	\caption{(Color online) Electronic structures of the lowest-enthalpy structures of MgSi$_2$ at, from top to bottom, 0 GPa ($R\overline{3}m$), 10 GPa ($Imma$), 20 GPa ($P6/mmm$), and 30 GPa ($P6/mmm$). The Fermi level (blue line) is placed at zero.} \label{fig:ebandsMgSi2}
\end{center}
\end{figure}

Although MgSi$_2$ is not thermodynamically stable at 0 GPa, it becomes stable at $P\simeq 10$ GPa and above. For completeness, we studied the lowest-lying structures of this compound at 0 GPa ($R\overline{3}m$), 10 GPa ($Imma$), 20 GPa and 30 GPa (both $P_6/mmm$). These structures, which are visualized in Fig. \ref{fig:strMgSi2}, are all dynamically stable, as demonstrated by the computed phonon band structures shown in Supplemental Material.\cite{supplement} Their computed electronic structures are given in Fig. \ref{fig:ebandsMgSi2}, showing that these phases are all metallic. At 0GPa, the $R\overline{3}m$ phase features profound local maximum of the density of electron states right above the Fermi level $E_{\rm F}$ while at 10 GPa, 20 GPa, and 30 GPa, such local maxima ($\simeq 3$ states/eV for $Imma$ phase at 10 GPa) are exactly at the Fermi level. The conduction bands contributing to these local maxima are primarily characterized by $\pi$-type bonding between adjacent Si atoms. They cross $E_{\rm F}$ several times, having multiple extremes and/or saddle points exactly at $E_{\rm F}$, leading to the van Hope singularities. Such ``flat band-steep band character'', which is a signature of possible superconductivity,\cite{ANIEBACK:ANIE199717881} has been widely used\cite{Alexey:CaB,C6CP08708F,ANGE:ANGE201701660} in the literature as a screening criterion when predicting superconducting structures of solids.

\begin{figure}[t]
\begin{center}
\includegraphics[width= 8.25 cm]{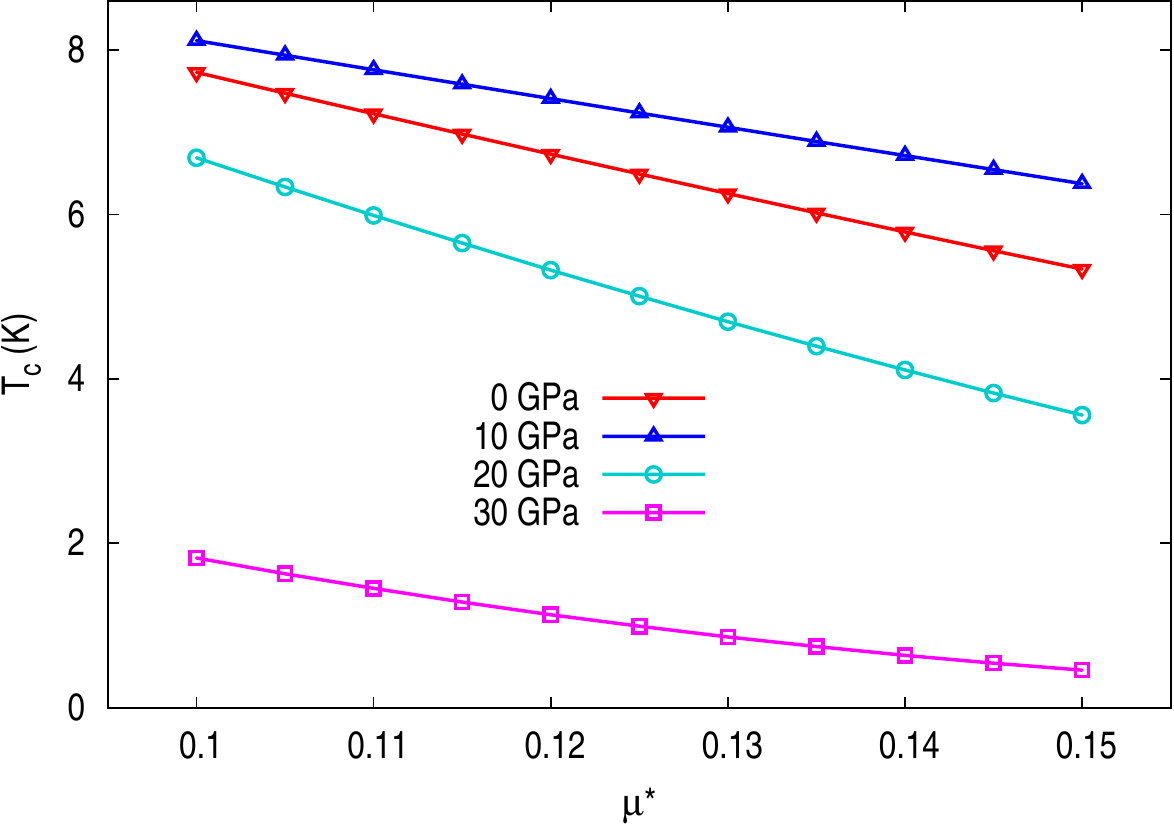}
\caption{(Color online) Superconducting critical temperature calculated for MgSi$_2$ at 0 GPa ($R\overline{3}m$ phase), 10 GPa ($Imma$ phase), and 20 GPa and 30 GPa  (both $P6/mmm$ phase).} \label{fig:Tc}
\end{center}
\end{figure}

The phonon-mediated superconducting properties of these predicted structures were computed with {\sc abinit} package,\cite{Gonze_Abinit_3} employing the linear response approach\cite{Abinit_phonon_2,Abinit_phonon_3} In short, we estimated the critical temperature $T_{\rm c}$ using the Allan-Dynes modified McMillan's approximation of the Eliashberg equation according to \cite{McMillanTc, AllenTc}
\begin{equation}\label{eq:Tc}
T_{\rm c} = \frac{\langle \omega_{\rm log}\rangle}{1.2}\exp\left[-\frac{1.04(1+\lambda)}{\lambda-\mu^*(1+0.62\lambda)}\right].
\end{equation}
Here, $\lambda$ is the overall electron-phonon coupling strength that can be computed from the frequency-dependent Eliashberg spectral function, $\langle \omega_{\rm log}\rangle$ the logarithmic average phonon frequency, and $\mu^*$ the Coulomb pseudopotential, for which the typical range of value (from $0.10$ to $0.15$)\cite{McMillanTc} was explored. For the $R\overline{3}m$ and $P6/mmm$ structures (3 atoms per primitive cell), the $k$-point mesh was chosen to be $15\times 15\times 15$ while for the $Imma$ with larger primitive cell (6 atoms), the $k$-point mesh was $12\times 12\times 12$. For all of these phase, the $q$-point mesh was $3\times 3\times 3$. We found that the experimentally realized $Imma$ phase of MgSi$_2$ is superconducting at $T_{\rm c}\simeq 6.9$ K. The other structural phases of this compound, i.e., $R\overline{3}m$ and $P6/mmm$, also display superconducting characteristics at the critical temperatures $T_{\rm c}$ up to $\simeq 7$ K, as shown in Fig. \ref{fig:Tc}.

\subsection{MgSi}
The proposed $P4/mmm$ structure of MgSi is composed of alternating Mg and Si rows along [100] direction.\cite{Matsuda1998,Ravi20044213} At 0 GPa and 10 GPa, it is higher than the $P2_1/m$ and $C2/m$ structures predicted herein by $\simeq 140$ meV/atom and $\simeq 38$ meV/atom, respectively. At higher pressure, the $P4/mmm$ structure becomes lowest in energy compared to the other structures of MgSi. This phase is about $0.1$ meV/atom above the convex hull at $20$ GPa while at $30$ GPa, it is thermodynamically stable. Crystallographic information, visualizations, and calculated phonon structures of these phases are given in Supplemental Materials, showing that they are distinct and dynamically stable.\cite{supplement}

\begin{figure}[t]
\begin{center}
\includegraphics[width= 8.25 cm]{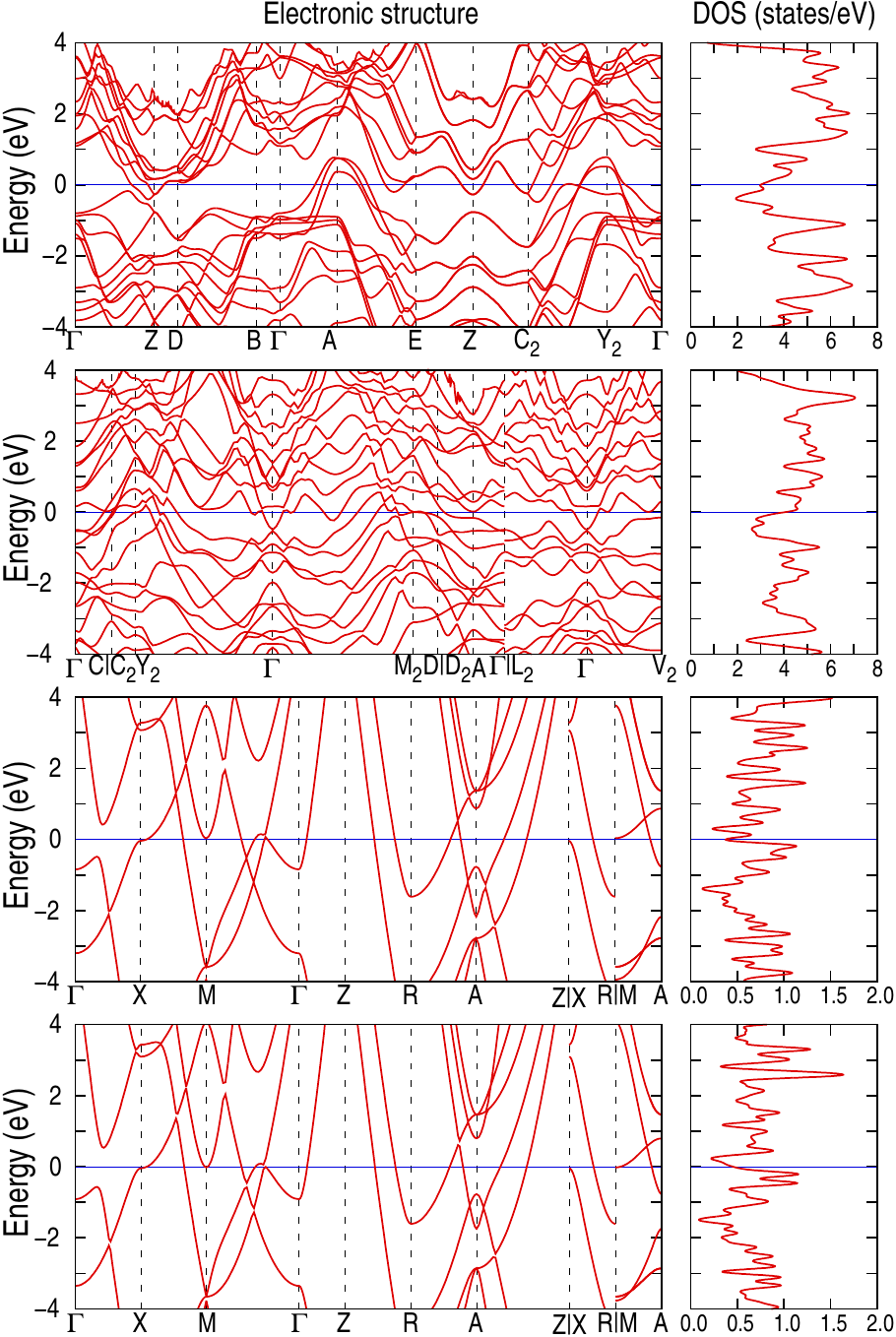}
	\caption{(Color online) Electronic structures of the lowest-enthalpy structures of MgSi at, from top to bottom, 0 GPa ($P2_1/m$), 10 GPa ($C2/m$), 20 GPa ($P4/mmm$), and 30 GPa ($P4/mmm$). The Fermi level (blue line) is chosen to be zero.} \label{fig:ebandsMgSi}
\end{center}
\end{figure}

All of these phases are metallic, as revealed in Fig. \ref{fig:ebandsMgSi} for their calculated electronic structures. At 0 GPa and 10 GPa, the $P2_1/m$ and $C2/m$ structures feature quite high density of electron states at the Fermi level $E_{\rm F}$, i.e., $\simeq 2.0$ and $\simeq 4.0$ states/eV, respectively. The density of electron states at $E_{\rm F}$ of the $P4/mmm$ structure at 20 GPa and 30 GPa is lower, roughly 0.5 states/eV. Due to the large primitive cells (12 atoms) and low symmetries (monoclinic) of the $P2_1/m$ and $C2/m$ structures, calculations for their $T_{\rm c}$ are substantially heavy, and for this reason, we have not done this work. However, it is possible that these phases are superconducting at 0 and 10 GPa with relatively high critical temperatures, possibly about $5$ K.

\section{Conclusions}
The family of Mg-Si binary compounds is rich of energetically competing phases, which are easily reordered by external pressure. This work provides some insights into the experimental observations of MgSi$_2$, MgSi, Mg$_9$Si$_5$, Mg$_5$Si$_6$, and Mg$_4$Si$_7$, whose proposed structures were found (computationally) to be thermodynamically unstable at ambient conditions. We find that at some finite pressures, the low-energy structures of MgSi$_2$, MgSi, and Mg$_9$Si$_5$ become stable. This result suggests that these binary compounds may likely be pressure-stabilized materials. The other two binary compounds considered, i.e., Mg$_5$Si$_6$ and Mg$_4$Si$_7$ are found to be unstable at pressure up to 30 GPa. This work supports previous experimental and computational studies,\cite{HastingJAP106, EHLERS2014617, qiu_kong_xiao_du_2016, Ninive2014, FLAMENT201764} claiming that the $\beta''$ phase of the Al-Mg-Si alloys is not Mg$_5$Si$_6$ as initially proposed. Similarly, the conjectured presence of Mg$_4$Si$_7$\cite{vanHuis20062945, VANHUIS20072183} remains an open question. On the other hand, some other compounds, including Mg$_2$Si$_3$ and Mg$_3$Si$_2$, are ``nearly'' stable at some ranges of pressure, and thus they may exist. Apparently, pressure should be considered for any computational studies of the formation of Mg-Si based solids, especially those found in their metastable phases. Finally, we find that MgSi$_2$ is a potential superconductor within a wide range of pressure with the critical temperature of the order of $5$ K.

\section*{Acknowledgements}
The author acknowledges Stefan Goedecker and Max Amsler for the minima-hopping code, Alexey Kolmogorov and Roxana Margine for useful discussion, and XSEDE for computational support through grant number TG-DMR170031. Valuable suggestions of two anonymous reviewers, which helped to improve the paper, are acknowledged.

%

\end{document}